\documentclass[12pt]{iopart}

\newcommand{\nn}{\nonumber}
\newcommand{\be}{\begin{equation}}
\newcommand{\ee}{\end{equation}}
\newcommand{\bea}{\begin{eqnarray}}
\newcommand{\eea}{\end{eqnarray}}

\usepackage{epsfig}
\usepackage{graphicx}
\usepackage{graphics}

\begin{document}

\title[Author guidelines for IOP journals in  \LaTeXe]{Decoherence of macroscopic states at finite temperatures}

\author{Carlos L. Benavides (1) and Claudia M. Ojeda (2)}

\address{(1) Departamento de F\'isica, Universidad de Los Andes, Bogot\'a D.C., Colombia}
\address{(2) Ecole Doctorale de Physique de la R\'egion Parisienne, Universit\'e Paris XI, Orsay, France}
\ead{ca-benav@uniandes.edu.co}

\begin{abstract}

We study the macroscopic superposition of light coherent states of the type Schr\"{o}dinger cat states; analizying, in particular, the role of the temperature in the decoherence processes, characteristic of the superposition of macroscopic states. The method we use here is based on the Master equation formalism, introducing an original approach. We use a modified Mandel function that is well adapted to the problem. This work is motivated by the experiments proposed by S. Haroche and collaborators in the 90's. In these experiments two Rydberg atoms were sent to a cavity in which a coherent state had been previously injected, monitoring the decay of quantum states due to dissipation. We find Haroche and collaborator's result at zero temperature and we predict the behavior of the field states in the cavity at finite temperatures.

\end{abstract}

\maketitle

\section{Introduction}
In classical physics, we cannot speak about superpositions of macroscopic states. Nevertheless, in quantum physics it is necessary to describe a system by a state vector which may be written as a coherent superposition of the eigenstates of some relevant observable. Roughly speaking, this is the Copenhagen interpretation of quantum mechanics. But the question is: what is relationship between the classical world and the microscopic description of physical reality?

The well known Schr\"{o}dinger's cat paradox suggests that if we assume that the rules of quantum mechanics are valid up the macroscopic level, then we have to conclude that it is possible to observe a superposition phenomena between distinguishable macroscopic states. In principle, this is only a theoretical point of view, but recently some groups have successfully achieved to produce this type of states in the laboratory. In a set of experiments performed in the 90's, S. Haroche and collaborators proposed an arrangement composed by a cavity prepared in a coherent state, two Ramsey zones and circular Rydberg atoms \cite{Haroche}. Using this configuaration they could create a coherent superposition of coherent states in the cavity. Their main goal was to study the dissipation of this type of state due to the irreversible coupling to the outside world reservoir.     
  
In order to predict the decoherence phenomena in the cavity, Haroche's group solved the master equation at zero temperature. They observed the transformation of the coherent superposition into a statistical mixture of orthogonal coherent states, and they found that for long times the state in the cavity evolves to a vacuum state. Our purpose is to study the effect of the temperature in the decoherence process in the cavity. In order to make this, we propose a modified Mandel function, solving the master equation for any temperature. In the first part of the paper, we offer a brief description of the experiment. In the second, we analyze the first part of the experiment and the steps to prepare a coherent state of light and then a superposition of these states. In the last part we study the decoherence process at finite temperatures for the coherent superposition of coherent states.  

\section{Brief description of the experiment}
\begin{figure}
    \centering
    \includegraphics[width=0.8\textwidth]{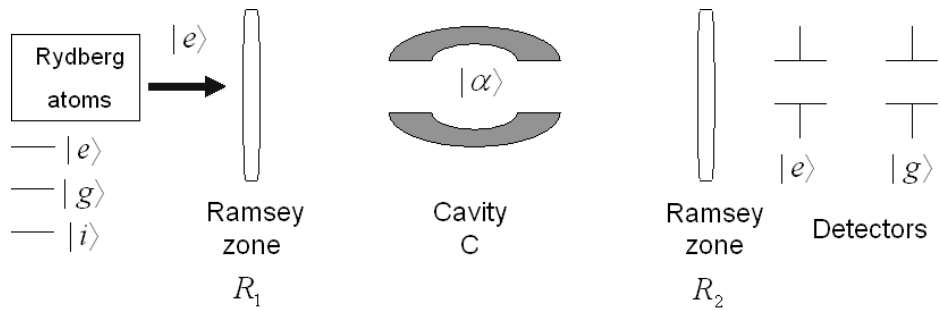}
    \caption{Setup proposed by Haroche {\it et. al.} to generate cat states in a cavity using Rydberg atoms.}
    \label{Experimento}
\end{figure}

We start by presenting the experimental setup proposed by Haroche and collaborators to prepare and detect coherent superpositions of classical states \cite{Haroche}. The experiment prepares a coherent superposition of classically distinct states of the electromagnetic field, a cat state, through the interaction of Rydberg atoms. The setup involves a cavity C prepared in a coherent state $|\alpha\rangle$, Rydberg atoms in resonance with cavity C and two additionals cavities $R_{1}$ and $R_{2}$ arranged as in the usual Ramsey method of interferometry. As it is sketched in figure \eref{Experimento}, an oven prepares the velocity selected circular Rydberg atoms. These atoms have principal quantum number $n\geq 30$ from which only three levels $|g\rangle$, $|e\rangle$ and $|f\rangle$ are required (n = 50,51,52). Circular levels are preferred because they are strongly coupled to microwaves and they have very long radiative decay times, which make them appropiate for preparing and detecting long-lived correlations between atoms and field states.

Before and after passing through cavity C the atoms go through cavities $R_{1}$ and $R_{2}$ respectively. These cavities are called Ramsey zones and they are in resonance with levels $|e\rangle$ and $|g\rangle$, whose transition frequency is $51.099 GHz$. They produce $\pi/2$ pulses over the atoms. Cavity C is placed between cavities $R_{1}$ and $R_{2}$. This cavity is made of superconducting niobium cooled to a temperature of about 1K and its dimensions are of $10^{-2}$ m. It supports a single mode of the quantized field of frequency $\nu_{C}=50GHz$. Cavity's frequency is tuned close to resonance with a transition connecting levels $|e\rangle$ and $|f\rangle$. This transition has a frequency of $48.180 GHz$ and is far from resonance of transitions involving level $|g\rangle$. The cavity mode is prepared in such a way that the field changes slowly along the atomic trajectory, which makes the atom field evolution adiabatic for slowly enough atoms and for sufficiently large detunings. At the end of the arragement atoms pass through two ionization zones $D_{e}$ and $D_{g}$. In these zones electric fields are applied to the atoms producing atomic ionization. They detect whether the atoms are in level $|e\rangle$ or in level $|g\rangle$ after they have crossed all the setup.

\section{Analysis of the first stage of the experiment. Preparation of a coherent superposition of two atoms, a cat state.}
We will present in this section the first stage of Haroche's experiment \cite{Haroche}, which consists on preparing an entangled state between the cavity field and the Rydberg atoms. It was this sort of state that Schr$\ddot{o}$dinger had in mind in 1935, in the sense that the macroscopic state of the cat is correlated with the microscopic state of the atom \cite{Knight}. Each atom coming from the oven is laser excited to state $|e\rangle$. After leaving the first Ramsey zone $R_{1}$ it is in a superposition of the two circular Rydberg states $|e\rangle$ and $|g\rangle$, such that 
\be\label{eR1}
|\psi_{atom}\rangle=\frac{1}{\sqrt{2}}\Big(|e\rangle+|g\rangle\big).
\ee

The interaction of the atom with the cavity is modeled by the interaction Hamiltonian,
\be
\hat{H_{I}}=\hbar\xi a^{\dagger} a\sigma_{3},
\ee
where $\sigma_{3}=|f\rangle\langle f|-|e\rangle\langle e|$, $\xi=2d^{2}/\Delta$, d is the atomic dipole moment and $\Delta$ the detuning. Before entering to the cavity the system is in the state
\be
|\psi_{atom-field}(0)\rangle=|\psi_{atom}\rangle|\alpha\rangle.
\ee
The atomic velocity is selected such that the atom leaves the cavity at $t=\pi/\xi$. At this moment the system is in the state
\bea\nn
|\psi_{atom-field}\rangle&=&e^{-i\hat{H_{I}}t/\hbar}|\psi_{field}(0)\rangle\\
\nn&=&\frac{1}{\sqrt{2}}\Big(|\alpha e^{i\xi t}\rangle|e\rangle+|\alpha\rangle|g\rangle\big)\\
\label{atomo-campo}&=&\frac{1}{\sqrt{2}}\Big(|e;-\alpha\rangle+|g;\alpha\rangle\big),
\eea
Finally, the atom is submitted to the second Ramsey zone $R_{2}$ after which state \ref{atomo-campo} becomes
\be\label{atomo-campo-R}
|\psi_{atom-field}\rangle=\frac{1}{2}\Big(|e;-\alpha\rangle-|e;\alpha\rangle+|g;\alpha\rangle+|g;-\alpha\rangle\Big).
\ee
The ionization zones $D_{e}$ and $D_{g}$ detects the state of the atom and the field in the cavity is collapsed to the following state,
\be
|\psi_{field}\rangle=\frac{1}{\sqrt{2\big(1+\cos\psi_{1}\exp(-2a^2)\big)}}\Big(|\alpha\rangle+e^{i\psi_{1}}|-\alpha\rangle\Big);
\ee
with $\psi_{1}=0$ if the atom is detected in state g and $\psi_{1}=1$ if the atom is detected in state e. This is a cat state, it describes the entanglement between the cavity and the atoms.

\section{Analysis of the second stage of the experiment. Monitoring the decoherence at finite temperature using a third atom, sending a mouse to the cavity.}
In the second stage of Haroche's experiment \cite{Haroche} a second atom is sent to the cavity in order to analize the decoherence process inside the cavity. We will analize the conditional probability of detecting the second atom in a certain state having measured the state of the first atom. The analysis of this probability as a function of the delay T between the two atoms will give us an idea of the evolution of the decoherence process in the cavity. As a contribution to the results obtained by Haroche and collaborators, we will introduce finite temperatures.

The second atom is sent a time T after the first atom. This time is assumed to be very small compared to the relaxation time scale of the cavity. The system's density operator after the atom has passed through the first Ramsey zone and before entering to the cavity is the following
\be\label{Atomo2CampoAntes}
\rho_{atom2-field}(T)=\frac{1}{2}\big(|e\rangle+|g\rangle\big)\big(|e\rangle+|g\rangle\big)\otimes\rho_{campo}(T).
\ee
After passing through the two Ramsey zones and the cavity, the density operator of the system can be written as \cite{Haroche}
\bea\label{Atomo2Campo}
\rho_{atom2-field}&=&\frac{1}{4}\Big[\big(|e\rangle+|g\rangle\big)\big(\langle e|+\langle g|\big)e^{-i\pi a^{\dagger}a}\rho_{field}(T)e^{i\pi a^{\dagger}a}\\
\nn &+&\big(-|e\rangle+|g\rangle\big)\big(-\langle e|+\langle g|\big)\rho_{field}(T)\\
\nn &+&\big(|e\rangle+|g\rangle\big)\big(-\langle e|+\langle g|\big)e^{-i\pi a^{\dagger}a}\rho_{field}(T)\\
\nn&+&\big(-|e\rangle+|g\rangle\big)\big(\langle e|+\langle g|\big)\rho_{field}(T)e^{i\pi a^{\dagger}a}\Big].
\eea
The phase operator $e^{-i\pi a^{\dagger}a}$ was introduced in state $|e\rangle$ since the cavity Hamiltonian affects only this state. From expression \eref{Atomo2Campo} it can be deduced that the probability of detecting the second atom in state n=e or g is given by the equation
\bea\label{Prob}
P_{n}&=&\langle n|Tr_{field}\Big(\rho_{atom2-field}\Big)|n\rangle\\
\nn&=&\frac{1}{2}\Big(1\pm Re\big(Tr_{field}(e^{-i\pi a^{\dagger}a}\rho_{field}(T)\big)\Big),
\eea

In order to calculate this probability, we propose a modified Mandel function which simplifies the calculation of the desired expressions. This function has the form
\be
Q_{m}(\beta)=\langle\beta|\rho|-\beta\rangle.
\ee
As is well known the master equation at finite temperatures is the following
\be\label{Master}
\dot{\rho}=-\frac{\varsigma}{2}n\big(aa^{\dagger}\rho-2a^{\dagger}\rho a+\rho aa^{\dagger}\big)-\frac{\varsigma}{2}(n+1)\big(a^{\dagger}a\rho-2a\rho a^{\dagger}+\rho a^{\dagger}a\big).
\ee
Applying the definition of this new Mandel function to the master equation, we have terms in the form $\langle\beta|aa^{\dagger}\rho|-\beta\rangle$, $\langle\beta|a^{\dagger}\rho a|-\beta\rangle$, $\langle\beta|\rho aa^{\dagger}|-\beta\rangle$, $\langle\beta|a^{\dagger}a\rho|-\beta\rangle$ and $\langle\beta|a\rho a^{\dagger}|-\beta\rangle$ whose calculation will be shown in the appendix. It can be found that the modified Mandel function $Q_{m}(\beta,\beta^{*},t)$ satisfies the following equation
\bea\label{Q}
\dot{Q}_{m}&=& \varsigma n\Big(4Q_{m}+8\left|\beta\right|^{2}Q_{m}+4\beta^{*}\frac{\partial}{\partial\beta^{*}}Q_{m}+4\beta\frac{\partial}{\partial\beta}Q_{m}+2\frac{\partial}{\partial\beta\partial\beta^{*}}Q_{m}\Big) \nn \\ &&+
\frac{\varsigma}{2}\Big(2Q_{m}+4\left|\beta\right|^{2}Q_{m}+3\beta^{*}\frac{\partial}{\partial\beta^{*}}Q_{m}+3\beta\frac{\partial}{\partial\beta}Q_{m}+2\frac{\partial}{\partial\beta\partial\beta^{*}}Q_{m}\Big).
\eea

We solve this equation for a general formulation of the density operator,
\bea\label{matrizdensidad}
\rho=A(t)\left|\alpha(t)\right\rangle\left\langle\gamma(t) \right|,
\eea
and we find as solution
\be\label{Qm}
\fl Q_{m}(\beta,\beta^{*},t) = A(t)\exp\Big(-\left|\beta\right|^{2}-\frac{\left|\alpha(t)\right|^{2}}{2}-\frac{\left|\gamma(t)\right|^{2}}{2}+\beta^{*}\alpha(t)-\beta\gamma(t)^{*}+\left|\beta\right|^{2}C(t)\Big),
\ee
where $A(t)$ and $C(t)$ are explicitly time depending functions. When we substitute this function $Q_{m}$ in equation \eref{Q} we find the following equations
\bea
\dot{C}(t) &=& -\varsigma n (C(t)+1)^{2}-\varsigma C(t)(C(t)+1), \\
\label{AA}\frac{\dot{A}(t)}{A(t)} &=& \frac{\dot{\alpha}\alpha^{*}}{2}+\frac{\dot{\alpha}^{*}\alpha}{2}+\frac{\dot{\gamma}\gamma^{*}}{2}+\frac{\dot{\gamma}^{*}\gamma}{2} +\varsigma(n+1)\gamma^{*}\alpha \nn \\ &&-\varsigma n (C(t)+1) - \varsigma C(t),\\
\label{alph}\dot{\alpha}(t) &=& -\varsigma n (C(t)+1)\alpha - \frac{\varsigma}{2}(1+2C(t))\alpha, \\
\label{gamm}\dot{\gamma}(t) &=& -\varsigma n (C(t)+1)\gamma - \frac{\varsigma}{2}(1+2C(t))\gamma.
\eea

The equation for $C(t)$ is the Ricatti equation, which can be reduced to the Bernoulli equation and then to a first order differential equation. The solution is the following:
\bea
C(t) = \frac{n e^{\varsigma t}}{(n+1)+n(n+1)(e^{\varsigma t}-1)}-\frac{n}{n+1}.
\eea
Solutions to equations \eref{AA}, \eref{alph} and \eref{gamm} are found to be:
\bea
\fl\alpha(t) &=& \alpha_{o}\frac{e^{-\frac{\varsigma}{2} t}}{1+n(e^{\varsigma t}-1)}, \\
\fl\gamma(t) &=& \gamma_{o}\frac{e^{-\frac{\varsigma}{2} t}}{1+n(e^{\varsigma t}-1)}, \\
\fl\label{A}A(t)&=&\exp\Big[\Big(-\frac{\alpha^{2}+\gamma^{2}}{2}+(n+1)\gamma\alpha\Big)R(n,t)+\big(\alpha^{2}+\gamma^{2}\big)nS(n,t)-nT(n,t)\Big],
\eea
where
\bea
R(n,t)&=&\int_{0}^{t}\frac{e^{-\varsigma t} dt}{\big(1+n(e^{\varsigma t}-1)\big)^{2}},\\
S(n,t)&=&\int_{0}^{t}\frac{dt}{\big(1+n(e^{\varsigma t}-1)\big)^{3}}
\eea
and
\be
T(n,t)=\int_{0}^{t}\frac{e^{\varsigma t} dt}{\big(1+n(e^{\varsigma t}-1)\big)}.
\ee

In order to calculate the conditional probability \eref{Prob}, the following trace must be calculated:
\bea
\Tr\Big(e^{-i\pi a^{\dagger}a}\rho_{field}(T)\Big)&=& Tr\Big(\rho_{field}(T)e^{-i\pi a^{\dagger}a}\Big) \\
\nn&=& \int\left\langle \beta\right|\rho_{field}(T)e^{-i\pi a^{\dagger}a}\left|\beta\right\rangle d^{2}\beta\\
\nn&=& \int\left\langle \beta\right|\rho_{field}(T)\left|-\beta\right\rangle d^{2}\beta,
\eea
where we can identify the modified Mandel function $Q_{m}$. If we consider a density operator of the form \eref{matrizdensidad} it suffices to integrate expression \eref{Qm}, getting as result
\bea
\nn \Tr\Big(e^{-i\pi a^{\dagger}a}\rho_{field}(T)\Big)&=& A(t)e^{-\frac{|\alpha(t)|^{2}}{2}}e^{-\frac{|\gamma(t)|^{2}}{2}}\int\exp\Big(-(1-C)\left|\beta\right|^{2}\\ \nn &&+\beta^{*}\alpha(t)-\beta\gamma(t)^{*}\Big) d^{2}\beta\\
\label{Tr}&=&A(t)e^{-\frac{|\alpha(t)|^{2}}{2}}e^{-\frac{|\gamma(t)|^{2}}{2}}\Big(\frac{1}{1-C}\Big)\exp\Big(\frac{-\alpha\gamma^{*}}{1-C}\Big).
\eea
At $t=0$, before the second atom has been sent to the cavity, the density operator associated to the cavity field had the form,
\be\label{RhoCampoAtomo1}
\rho_{field}=\frac{1}{N_{1}^{2}}\big(|\alpha\rangle\langle\alpha|+|-\alpha\rangle\langle-\alpha|+e^{i\psi_{1}}|-\alpha\rangle\langle\alpha|+e^{-i\psi_{1}}|\alpha\rangle\langle-\alpha|\big).
\ee
where $N_{1}=\sqrt{2\big(1+\cos\psi_{1}\exp(-2a^{2})\big)}$. To get the probability of having the second atom in state e or g, we proceed to make the same calculation as in \eref{Tr} for each of the terms in expression \eref{RhoCampoAtomo1}. For the terms $|\alpha\rangle\langle\alpha|$ and $|-\alpha\rangle\langle-\alpha|$, we get
\be
\Tr\Big(e^{-i\pi a^{\dagger}a}|\alpha\rangle\langle\alpha|\Big)=\Tr\Big(e^{-i\pi a^{\dagger}a}|-\alpha\rangle\langle-\alpha|\Big)=A_{1}(t)e^{-|\alpha(t)|^{2}\big(\frac{2-C}{1-C}\big)}\Big(\frac{1}{1-C}\Big)
\ee
and for the terms $|\alpha\rangle\langle-\alpha|$ and $|-\alpha\rangle\langle\alpha|$ we obtain
\be\nn
\Tr\Big(e^{-i\pi a^{\dagger}a}|\alpha\rangle\langle-\alpha|\Big)=\Tr\Big(e^{-i\pi a^{\dagger}a}|-\alpha\rangle\langle\alpha|\Big)=A_{2}(t)e^{|\alpha(t)|^{2}\big(\frac{C}{1-C}\big)}\Big(\frac{1}{1-C}\Big)
\ee

Replacing these results into equation \eref{Prob}, we have the conditional probability time evolution of finding the second atom in state n=e or g given the first atom was sent to the cavity in state e,
\be\label{ProbFin}
P(g,e;T)=\frac{1}{2}\Bigg\{1\pm Re\Bigg[\frac{2}{N_{1}^{2}}\Big(\frac{1}{1-C}\Big)\Big(A_{1}(t)e^{-|\alpha(t)|^{2}\big(\frac{2-C}{1-C}\big)}+\cos\psi_{1}A_{2}(t)e^{|\alpha(t)|^{2}\big(\frac{C}{1-C}\big)}\Big)\Bigg]\Bigg\},
\ee
where $A_{1}(t)$ is the expression for $A(t)$ making $\alpha_{0} = \gamma_{0}$, and $A_{2}(t)$ is the expression for $A(t)$ making $\alpha_{0} = -\gamma_{0}$.

\begin{figure}[h]
    \centering
    \includegraphics[width=0.6\textwidth]{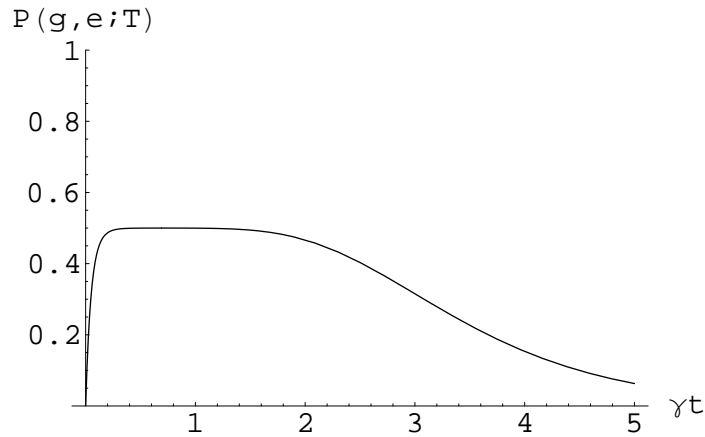}
    \caption{Conditional probability P(g,e;T) of detecting the second atom in level e after having detected the first one in level g, as a function of the delay T between the two atoms, for the experiment sketched in Fig. 1. The average number of photons in the cavity is equal to 10 at zero temperature.}
    \label{nigual0}
\end{figure}

\begin{figure}[h]
    \centering
    \includegraphics[width=0.6\textwidth]{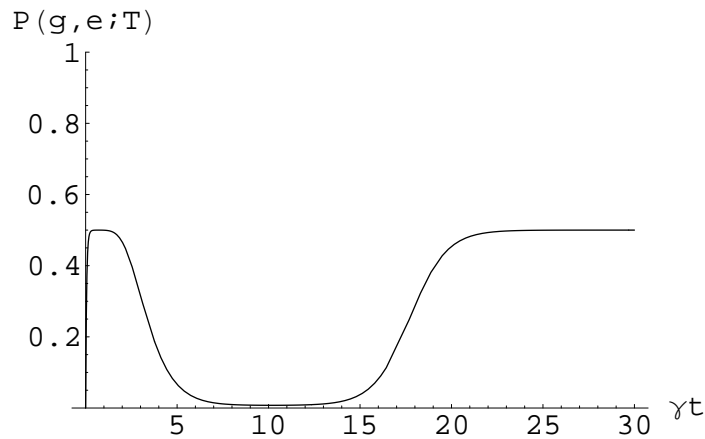}
    \caption{Conditional probability P(g,e;T) of detecting the second atom in level e. The average number of photons in the cavity is equal to 10 at finite temperature (n = 0.00001).}
    \label{npeque}
\end{figure}

\begin{figure}[h]
    \centering
    \includegraphics[width=0.6\textwidth]{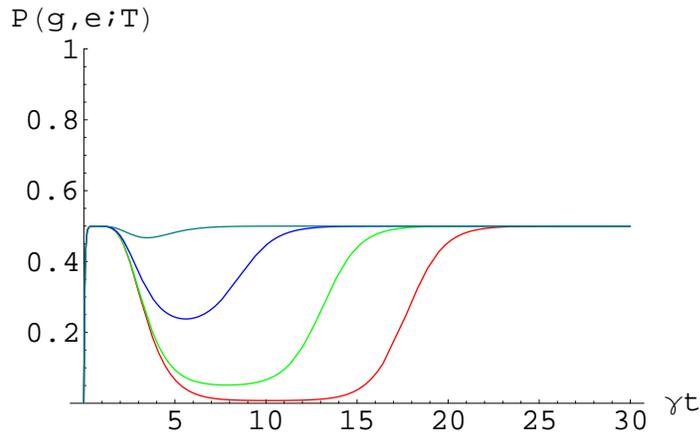}
    \caption{Conditional probability P(g,e;T) of detecting the second atom in level e. The average number of photons in the cavity is equal to 10 at finite temperatures (black line n = 0.1, blue line n= 0.001, green line n = 0.0001 and red line 0.00001).}
    \label{nvarioa}
\end{figure}

\section{Discussion}
In this paper we have investigated the role of the temperature in the macroscopic superposition of orthogonal coherent states of light. The main motivation of our work is based on the fact that at finite temperatures there are non trivial effects in the entanglement of quantum sates. Introducing finite temperatures lead us to a more realistic approach to this phenomena. 

As it is sketched in fig. 2 and fig. 3, for short times, we see a rapidly evolution of the coherent superposition to a statistical mixture. This fast evolution is the signature of the rapid decoherence process between the two orthogonal states of the cavity. The plateau $P_{n} = 1/2$ reveals the appearance of a statistical mixture of light states. The following decay indicates the incoherent superposition of the two states and their overlap due to energy dissipation on the cavity \cite{Haroche}.

For values of the temperature different to zero, there is a revival of statistical mixture of the orthogonal states of light in the cavity. In these cases, the system evolves from a quasi-vacuum state to equilibrium statistical mixture. Surprisingly, we have shown that the temperature helps the system to maintain the superposition of macroscopic states. As it is sketched in fig. 4, the value of the temperature has a very near relation with the time between the two plateaus.  

\appendix
\section{Calculation of the modificated Mandel function}
The modified Mandel function has the form
\be
Q_{m}(\beta)=\langle\beta|\rho|-\beta\rangle.
\ee
In order to find the differential equation that satisfies this function, we apply the master equation for finite temperatures,
\be
\dot{\rho}=-\frac{\varsigma}{2}n\big(aa^{\dagger}\rho-2a^{\dagger}\rho a+\rho aa^{\dagger}\big)-\frac{\varsigma}{2}(n+1)\big(a^{\dagger}a\rho-2a\rho a^{\dagger}+\rho a^{\dagger}a\big).
\ee
Calculating separately each of the terms we have,
\bea\label{1}
\fl\langle\beta|aa^{\dagger}\rho|-\beta\rangle&=&\langle\beta|\rho|-\beta\rangle+\beta^{*}\Big[\Big(\frac{\partial}{\partial \beta^{*}}+\frac{\beta}{2}\Big)\langle\beta|\Big]\rho|-\beta\rangle\\
\fl\nn&=&Q_{m}+|\beta|^{2}Q_{m}+\beta^{*}\frac{\partial}{\partial\beta^{*}}Q_{m}, \\
\fl\langle\beta|a^{\dagger}\rho a|-\beta\rangle&=&-|\beta|^{2}Q_{m},\\
\fl\langle\beta|\rho aa^{\dagger}|-\beta\rangle&=&\langle\beta|\rho|-\beta\rangle+\frac{|\beta|^{2}}{2}\langle\beta|\rho|-\beta\rangle+\beta\Big[\frac{\partial}{\partial \beta}\langle\beta|\rho|-\beta\rangle-\Big(\frac{\partial}{\partial\beta}\langle\beta|\Big)\rho|-\beta\rangle\Big]\nn\\
\fl&=&Q_{m}+|\beta|^{2}Q_{m}+\beta\frac{\partial}{\partial\beta}Q_{m},\\
\fl\langle\beta|a^{\dagger}a\rho|-\beta\rangle&=&\beta^{*}\Big[\Big(\frac{\partial}{\partial \beta^{*}}+\frac{\beta}{2}\Big)\langle\beta|\Big]\rho|-\beta\rangle\nn\\
\fl&=&|\beta|^{2}Q_{m}+\beta^{*}\frac{\partial}{\partial\beta^{*}}Q_{m}, \\
\fl\langle\beta|a\rho a^{\dagger}|-\beta\rangle&=&-\Big[\Big(\frac{\beta}{2}+\frac{\partial}{\partial\beta^{*}}\Big)\langle\beta|\Big]\rho\Big[\Big(\frac{\partial}{\partial\beta}+\frac{\beta^{*}}{2}\Big)|-\beta\rangle\Big]\\
\fl\nn&=&-\frac{|\beta|^{2}}{4}\langle\beta|\rho|-\beta\rangle-\frac{\beta}{2}\langle\beta|\rho\Big(\frac{\partial}{\partial\beta}|-\beta\rangle\Big)\\ \fl\nn&-&\frac{\beta^{*}}{2}\Big(\frac{\partial}{\partial\beta^{*}}\langle\beta|\Big)\rho|-\beta\rangle-\Big(\frac{\partial}{\partial\beta^{*}}\langle\beta|\Big)\rho\big(\frac{\partial}{\partial\beta}|-\beta\rangle\Big)\\
\fl\nn&=&-\frac{|\beta|^{2}}{4}\langle\beta|\rho|-\beta\rangle-\frac{\beta}{2}\Big[\frac{\partial}{\partial\beta}Q_{m}+\frac{\beta^{*}}{2}Q_{m}\Big]-\frac{\beta^{*}}{2}\Big[\frac{\partial}{\partial\beta^{*}}Q_{m}+\frac{\beta}{2}Q_{m}\Big]\\
\fl\nn&-&\Big[\frac{\partial}{\partial\beta\partial\beta^{*}}Q_{m}+Q_{m}+\frac{\beta^{*}}{2}\frac{\partial}{\partial\beta^{*}}Q_{m}+\frac{\beta}{2}\frac{\partial}{\partial\beta}Q_{m}+\frac{|\beta|^{2}}{4}Q_{m}\Big],
\eea
which conduces us to the differential equation of the modified Mandel function,
\bea\label{Q}
\dot{Q}&=& \varsigma n\Big(4Q_{m}+8\left|\beta\right|^{2}Q_{m}+4\beta^{*}\frac{\partial}{\partial\beta^{*}}Q_{m}+4\beta\frac{\partial}{\partial\beta}Q_{m}+2\frac{\partial}{\partial\beta\partial\beta^{*}}Q_{m}\Big) \nn \\ &&+
\frac{\varsigma}{2}\Big(2Q_{m}+4\left|\beta\right|^{2}Q_{m}+3\beta^{*}\frac{\partial}{\partial\beta^{*}}Q_{m}+3\beta\frac{\partial}{\partial\beta}Q_{m}+2\frac{\partial}{\partial\beta\partial\beta^{*}}Q_{m}\Big). \nn \\
\eea


\section*{References}
\begin{thebibliography}{10}
\bibitem{Haroche} Davidovich L, Brune M, Raimond J M and Haroche S 1996 {\it Phys. Rev. A} {\bf 53} 1295-1309.
\bibitem{Knight} Gerry C C and Knight P L 1997 {\it Am. J. Phys.} {\bf 65} 964-974.
\end{thebibliography}
\end{document}